\documentclass[twocolumn]{article}
\usepackage{epsfig}

\begin{document}

\boldmath
\title{\bf Evidence of the $\phi \to \eta \pi^0 \gamma$ decay}
\unboldmath

\normalsize
\author{
M.N.Achasov,
V.M.Aulchenko,
S.E.Baru,
A.V.Berdyugin, \\
A.V.Bozhenok,
A.D.Bukin, 
D.A.Bukin,
S.V.Burdin,
T.V.Dimova, \\
S.I.Dolinski, 
V.P.Druzhinin,
M.S.Dubrovin, 
I.A.Gaponenko, \\
V.B.Golubev\thanks{email:golubev@inp.nsk.su}, 
V.N.Ivanchenko,
P.M.Ivanov,
A.A.Korol, \\
S.V.Koshuba,
A.P.Lysenko,
E.V.Pakhtusova,
A.A.Salnikov, \\
S.I.Serednyakov,
V.V.Shary,
Yu.M.Shatunov,
V.A.Sidorov, \\
Z.K.Silagadze,
A.N.Skrinsky,
Yu.V.Usov
\\
\\
\date{}
{Budker Institute of Nuclear Physics, Novosibirsk, 630090, Russia}
}
\maketitle

\begin{abstract}

Signal of the rare radiative decay $\phi \to \eta \pi^0 \gamma$
was observed in the
SND experiment at VEPP-2M electron-positron collider.
The result is based on the analysis of data, corresponding to
a total integrated luminosity of $4 pb^{-1}$, or $8\cdot 10^{6}$
$\phi$-mesons produced. The measured branching ratio of 
$\phi \to \eta \pi^0 \gamma$ decay is equal to $(0.83\pm 0.23)\cdot 10^{-4}$.

\end{abstract}

\section{Introduction}

Several mechanisms may contribute into $\phi(1020) \to \eta \pi^0 \gamma$ decay
rate including  $\phi\to V\pi^0$ type
transitions, where $V$ is a vector meson, with
a subsequent $V\to\eta\gamma$ radiative decay, and electric dipole
radiative transition $\phi\to a_0(980)\gamma$ with a scalar $a_0(980)$ meson
decaying into $\eta\pi^0$.

Theoretical predictions of the  $\phi \to \eta \pi^0 \gamma$ branching
ratio, based on assumption of pure $\phi\to V\pi^0$ decay 
mechanism (\cite{TH1},
\cite{TH3}, \cite{TH4}), are quite low, of the order of $5\cdot10^{-6}$.
Situation with
theoretical description of the $\phi\to a_0(980)\gamma$ decay
is much more complicated, because the results strongly depend on the
quark structure of the $a_0(980)$ meson.
At present, its quark structure
is not well established and several models
exist including modifications of $q\bar{q}$ scheme \cite{TH5}, 
4-quark model \cite{TH6}, and $K\overline{K}$  molecular
model \cite{TH7}. Theoretical estimations of $\phi\to a_0(980)\gamma$
branching ratios vary from about $10^{-5}$ in two-quark and
$K\overline{K}$ 
molecular models up to $10^{-4}$ in the 4-quark model
(\cite{TH1},\cite{TH2},\cite{TH3}), making the  $\phi\to a_0(980)\gamma$
decay a unique probe of the structure of $a_0(980)$ meson.

    The first search for the $\phi\to\eta\pi^0\gamma$ decay was conducted
in the ND experiment \cite{ND} at VEPP-2M  $e^+e^-$ collider in 1987,
where an upper limit of $2.5\cdot 10^{-3}$ was established. In 1995 the new
SND detector having better hermeticity, granularity,
energy and spatial resolution started operation at VEPP-2M. First indications
of $\phi\to\eta\pi^0\gamma$ decay were seen by SND in 1997 \cite{HADR97}.
The result was based on analysis of half of the $\phi$-meson
data recorded in 1996--1997. In this work analysis of full data sample
was carried out. The
$\phi\to\eta\pi^0\gamma$ decay was observed for the first time and its
branching ratio was measured.   

\section{Detector and experiment}
Experiment  \cite{Prep.96}, \cite{Prep.97}
was carried out in 1996 at VEPP-2M $e^+e^-$ collider with SND
detector. The SND detector \cite{SND} is a universal nonmagnetic detector.
Its main part is a 3-layer electromagnetic calorimeter consisted of 1630
NaI(Tl) crystals with a total thickness of 35 cm or 14 radiation lengths. 
The energy resolution for photons can be described as
$4.2\%/\sqrt[{4\enskip} ]{E(GeV)}$ \cite{Calibr.}, the angular
resolution of the
calorimeter is close to 1.5 degrees, and the solid angle coverage
is $90\%$ of $4\pi$ steradian.

The data, used for the study of $\phi \to \eta\pi^0 \gamma$ decay was
recorded in 1996-1997 in the energy range from 980 to 1040 MeV
with the most of the data collected in the close vicinity of the $\phi$-peak.
Six successive scans of the energy range were performed with a total
integrated luminosity of $4.0 pb^{-1}$ and total number of $\phi$-mesons
produced of $8.2\cdot10^6$.
 
\section{Data analysis}

The main background sources for the process under study

\begin{equation}
e^+e^- \to \phi \to \eta\pi^0 \gamma \to 5\gamma
\end{equation}
are the following $\phi$-meson decays:
\begin{equation}
e^+e^- \to \phi \to \eta\gamma \to 3\pi^0\gamma
\end{equation}
\begin{equation}
e^+e^- \to \phi \to K_S K_L \to neutrals
\end{equation}
\begin{equation}
e^+e^- \to \phi \to \pi^0\pi^0\gamma
\end{equation}
and nonresonant process
\begin{equation}
e^+e^- \to \omega\pi^0 \to \pi^0\pi^0\gamma.
\end{equation}

The expected number of events of the process (1) at a branching
ratio of  $\phi \to \eta\pi^0 \gamma$ equal to $10^{-4}$ is about 300, while
the number of background events (2) is $3\cdot10^4$. Although the process (2)
does not produce $5\gamma$ events directly, it can fake the topology of the
process (1)  due to either merging of close photons or loss of
soft photons through openings in the calorimeter. The process (3) produces
$8\cdot10^5$ $K_SK_L$ events with $K_S\to\pi^0\pi^0$ decays.
The $K_L$-s in $\phi$-meson decays are slow and have a decay length of
3 m, while the nuclear interaction length in NaI(Tl) is about 30 cm.
Characteristic feature of spurious $5\gamma$ events produced by $K_SK_L$ decays
due to either nuclear interaction of $K_L$-s or their decays in flight
is an energy-momentum imbalance and poor quality of at least one
photon in the event.

Primary event selection was based on simple criteria: the number of
reconstructed photons is equal to five, there are no tracks in the central
drift chambers, the total energy deposition ranges from 0.8 to 1.1 center
of mass energy $2E_0$, and the total transverse momentum is less
than $0.15E_0/c$. Such criteria greatly reduce background from the
processes (2) and (3), not affecting the decays (1), (4), and (5).
Also rejected were the events containing close photon pairs with spatial 
angle between photons less than 27 degrees, where energies
could be poorly reconstructed.

Next step in the event selection was based on quality of photons and
kinematic fit.
Parameter $\zeta$ \cite{XINM} was constructed describing
the likelihood
of a hypothesis, that given transverse energy deposition
profile of a cluster of
hit crystals in the calorimeter can be attributed to a single photon. The
requirements were imposed that  $\zeta<0$, for all photons.
Then kinematic fit was performed under the assumption, that selected
events are $e^+e^-\to 5\gamma$ ones and corresponding parameter $\chi_e^2$,
describing the likelihood of this assumption was calculated. Events with
$\chi_e^2>10$ were also rejected. Study of simulated $5\gamma$ events of
the processes (1), (4), and (5) showed, that the  $\chi_e^2$
and  $\zeta$ cuts reject less than $15\%$ of true $5\gamma$ events,
suppressing the process (2) by a factor of 3 and making the
expected background due to the process (3) very small, of the order of
10 events. Although, in contrast with the other processes, the Monte Carlo
simulation of the process (3) is much less accurate due to nuclear interaction
of $K_L$, so the data analysis may not completely rely on this estimation.

\begin{figure}[b!]
  \begin{center}
      \mbox{\epsfig{figure=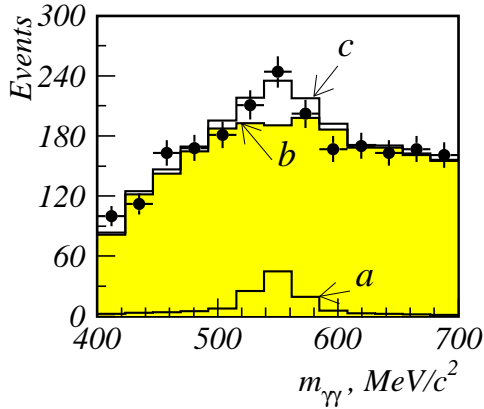,%
                              width=80 mm}}
  \end{center}
\caption{ Invariant masses of pairs of most energetic photons:
circles with error bars -- experimental data,
a -- simulated signal from $\phi\to\eta\pi^0\gamma$ decay corresponding to a
branching ratio of $0.7\cdot10^{-4}$, b -- estimated background
from the $e^+e^-\to\omega\pi^0$ and $\phi\to\eta\gamma,
f_0(980)\gamma$ events, c -- sum of a and b}
  \label{INCLUSIVE}
  \end{figure}

Characteristic feature of the
process (1) is that each event must contain two photon pairs with invariant
masses of $\eta$ and $\pi^0$ mesons. Simulation shows that if the
photons in an event are enumerated
in descending order in energy, the photons from $\eta\to\gamma\gamma$ decay
have the numbers of either 1 and 2 or 1 and 3.
Corresponding experimental and simulated
distributions in $m_{12}$ and $m_{13}$ with an additional
requirement, that rest of photons contains a pair with
$|m_{ij}-m_{\pi^0}|<20~MeV/c^2$ and more stringent photon quality requirement
$\zeta<-5$.
are shown in Fig.\ref{INCLUSIVE}. Background estimations are based on PDG table
value for $\phi\to\eta\gamma$ decay branching ratio \cite{PDG},
our measurements of $e^+e^-\to\omega\pi^0$ cross section
\cite{Prep.97} and $\phi\to\pi^0\pi^0\gamma$ decay branching ratio \cite{PPG}.
Distribution of experimental events
in $m_{12}+m_{13}$ shows an enhancement
centered at $\eta$-meson mass, while the sum of background
processes is nearly flat in this region. The sum of all
simulated background processes, where each one was normalized to the
number of events expected at a given integrated luminosity and total number
$\phi$-meson decays, describes the spectrum outside the
enhancement quite well. If this excess of events is due to the decay (1), it
corresponds to a $\phi\to\eta\pi^0\gamma$ branching ratio
of the order of $7\cdot10^{-5}$, but it is difficult
to extract accurate result from these inclusive
spectra, because of poor signal to background ratio and unknown decay
dynamics.

Detailed study of the process (1) requires substantial suppression of
background. It could be done using kinematic fit with intermediate $\eta$
and $\pi^0$ mesons
taken into account. For each event $\eta\pi^0\gamma$ and $\pi^0\pi^0\gamma$
hypotheses were tried and corresponding
$\chi_{\eta\pi^0\gamma}^2$,
and $\chi_{\pi^0\pi^0\gamma}^2$ parameters calculated.
To suppress the processes (4) and (5) the following requirements were imposed:
$\chi_{\eta\pi^0\gamma}^2<20$,
$\chi_{\pi^0\pi^0\gamma}^2>20$, 
and for additional suppression the process (2): $\zeta<-4$. With these
requirements the contribution from the processes (4) and (5), which are 
relatively rare themselves, becomes negligible.

\begin{figure}[b!]
  \begin{center}
      \mbox{\epsfig{figure=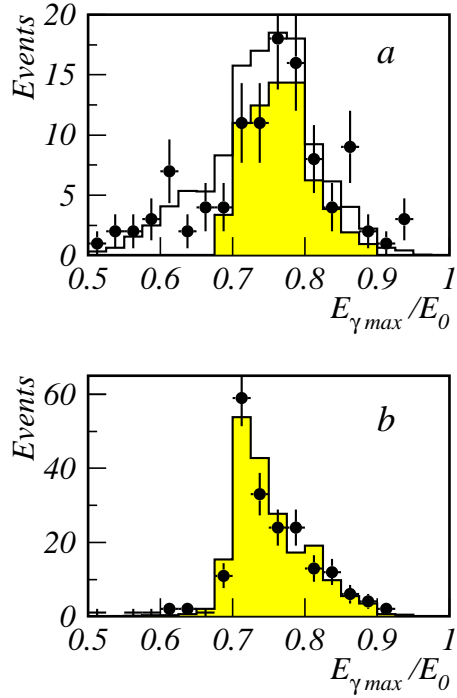,%
                              width=80mm}}
  \end{center}
\caption{ Spectra of the most energetic photon in the event:
a -- photons quality $\zeta<-4$, b -- $0<\zeta<10$. 
Circles with error bars depict experimental data,
shaded histogram --
 simulation of the process (2) and clear histogram -- simulated sum
of (2) and (1) with a $BR=10^{-4}$.}
  \label{ER1N}
  \end{figure}

The resulting spectra of the $E_{\gamma max}/E_0$, energy of the most
energetic photon, are shown in Fig.\ref{ER1N}.
Since the recoil photon in the process (2) has a narrow spectrum peaked at
360 MeV,
$E_{\gamma max}/E_0$ must be more than 0.7 in this process.
The spectrum in Fig.\ref{ER1N}b,
where true 5$\gamma$ events are suppressed by an order of magnitude,
is well reproduced by simulation of the process (2) alone, while the
spectrum in Fig.\ref{ER1N}a shows excess of experimental events 
with $\zeta<-4$ over the simulation of (2).
The background (2) still dominates at $E_{\gamma max}/E_0>0.7$, so
events with $E_{\gamma max}/E_0>0.7$ were rejected. From Fig.\ref{ER1N}a
one can see, that this roughly halves the detection efficiency for the process
(1). This also enhances dependence of the detection efficiency
on the invariant mass of $\eta\pi^0$ system, which varies from
$1\%$ at $M_{\eta\pi^0}=970~MeV/c^2$ up to
$5\%$ at $M_{\eta\pi^0}=700~MeV/c^2$.

\begin{figure}[b!]
  \begin{center}
      \mbox{\epsfig{figure=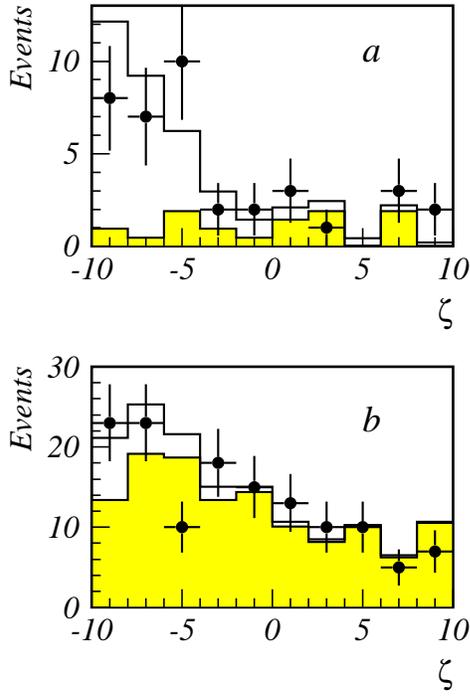,%
                              width=80mm}}
  \end{center}
\caption{ Distribution of events over $\zeta$:
a -- events with  $E_{\gamma max}/E_0<0.7$, b -- events
with  $0.7<E_{\gamma max}/E_0<0.8$
circles with error bars -- experimental data,
 shaded histogram
depicts simulation of the process (2) and clear histogram -- simulated sum
of (2) and (1) with a $BR=10^{-4}$.}
  \label{XINM}
  \end{figure}

The distributions of the events over photon quality parameter $\zeta$ are
shown in Fig.\ref{XINM}. Selection criteria here were the same as described 
above, except loosened
requirement on photon quality: $\zeta<10$. It can be seen,
that simulation well describes the background from the process (2), while
the excess of events in Fig.\ref{XINM}a at low $\zeta$ is compatible with
existence of the decay (1) with BR of the order of $10^{-4}$. Also was checked
the distribution in $\chi^2_{\eta\pi^0\gamma}$. 
While the distributions of experimental events with $0<\zeta<10$ is in 
a good agreement with simulation of the process (2), the distribution of the
events with $\zeta<-4$ agrees with the simulation of the process (1) with
a small contribution of the process (2). The total number of
selected events satisfying the selection criteria
is 25, from which 5 were estimated to be a background from the process (2).
The corresponding branching ratio was calculated using the $\eta\pi^0$
invariant mass distribution of the selected events and detection
efficiencies, obtained from simulation. The resulting value is
$(0.83\pm0.23)\cdot10^{-4}$, where error is a statistical one. For such
stringent selection criteria the systematic error in detection efficiency 
could be as high as 15\%, but at present level of experimental statistics
the statistical error dominates.

\begin{figure}
  \begin{center}
      \mbox{\epsfig{figure=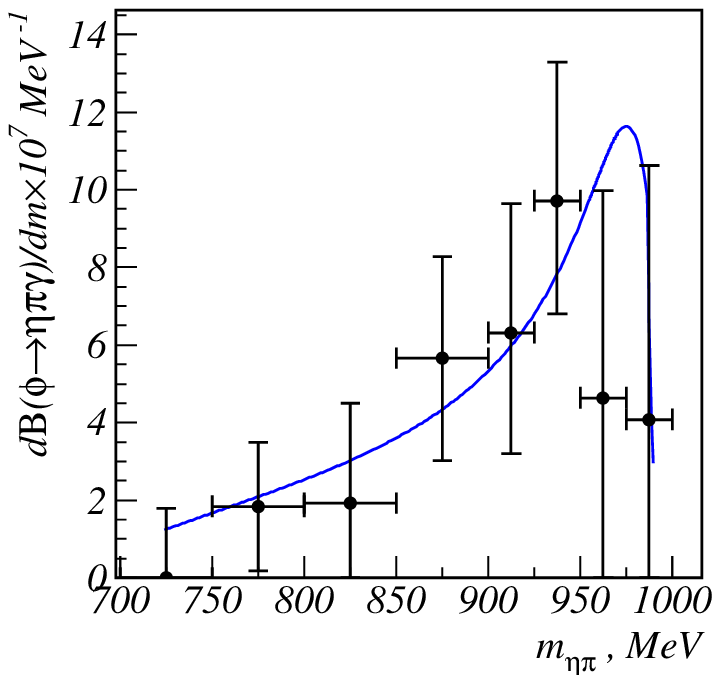,%
                              width=80mm}}
  \end{center}
\caption{ $\eta\pi^0$ invariant mass spectrum:
points  -- experimental data, curve -- optimal fit according to Ref.
\cite{TH1}.}
  \label{EPMASS}
  \end{figure}

In order to study the $\eta\pi^0$ mass spectrum the
$E_{\gamma max}$ cut was dropped. This increased the background from
the $\phi\to\eta\gamma$ decay, but made detection efficiency twice higher and
less dependent on $\eta\pi^0$ mass. The $\eta\gamma$ background was subtracted
using events with $M_{\gamma\gamma}$ outside $\eta$ peak. The result is shown
in Fig.\ref{EPMASS}. The invariant mass spectrum was fitted using formulae
from Ref.\cite{TH1} and assuming that the decay is a pure
 $\phi\to a_0(980)\gamma$ transition. The result is also shown in
Fig.\ref{EPMASS}. It is seen, that although the errors are very large,
the observed mass spectrum is consistent with
$a_0\gamma$ decay mechanism.
Constraint fit with $g_{a\eta\pi}=0.85g_{aK^+K^-}$ \cite{TH6}
gives the following values of $a_0(980)$ meson parameters: \\
$M_{a_0}=986^{+23}_{-10}~MeV/c^2$, \\
$g^2_{aK^+K^-}/4\pi=(1.5\pm 0.5)~GeV^2$.

\section{Conclusion} In this work evidence of the new decay
$\phi\to\eta\pi^0\gamma$ with a branching ratio of 
$(0.83\pm 0.23)\cdot 10^{-4}$ is presented. The observed enhancement
in the  $\eta\pi^0$ invariant mass spectrum at large  $m_{\eta\pi^0}$
is consistent with domination of $a_0\gamma$ decay mechanism, while
relatively high decay rate agrees with the theoretical predictions based
on 4-quark model of $a_0(980)$-meson \cite{TH1},
\cite{TH3}.

\section{Acknowledgements}
    This work is supported in part by
Russian Fund for Basic Researches, grants No. 96-02-19192 and 96-15-96327,
and STP ``Integration'' No.274.

%%%%%%%%%%%%% Figures %%%%%%%%%%%%%%%

%%%%%%%%%%%%% End of figures  %%%%%%%%%%%%%%%


\begin {thebibliography}{10}

\bibitem{TH1}
N.N.Achasov, V.N.Ivanchenko, Nucl. Phys., B 315 (1989) 465.

\bibitem{TH3}
N.N.Achasov, V.V.Gubin, Phys. Rev. D 56 (1997) 4084.

\bibitem{TH4}
A.Bramon, A.Grau and G.Pancheri, Phys. Lett. B 283 (1992) 416.

\bibitem{TH5}
N.T\"{o}rnqvist, Phys. Rev. Lett. 49 (1982) 624.

\bibitem{TH6}
R.L.Jaffe, Phys. Rev. D 15 (1977) 267, 281.

\bibitem{TH7}
J.Weinstein, N.Isgur, Phys. Rev. D 41 (1990) 2236.

\bibitem{TH2}
F.E.Close, N.Isgur, S.Kumano, Nucl. Phys. B 389 (1993) 513.

\bibitem{ND}
V.P.Druzhinin et al., Phys. Reports 202 (1991) 99.

\bibitem{HADR97} M.N.Achasov et al.,
e-print archive: hep-ex/9711023;
Proc. of HADRON97 Conference, Upton NY, August 24--30,
1997, p.26--35, p.783--786.
To be published in Phys. Atom. Nucl..

\bibitem{Prep.96}
M.N.Achasov et al., Preprint Budker INP 96-47, Novosibirsk, 1996.

\bibitem{Prep.97}
M.N.Achasov et al., Preprint Budker INP 97-78, Novosibirsk, 1997.

\bibitem{SND}
V.M.Aulchenko et al., Proc. Workshop on Physics and Detectors for DA$\Phi$NE,
Frascati, April 9-12, (1991), p.605.

\bibitem{Calibr.}
M.N.Achasov et al., Nucl. Instr. and Meth. A401(1997)179

\bibitem{XINM} A.V.Bozhenok, V.N.Ivanchenko, Z.K.Silagadze,
Nucl. Instr. and Meth. A 379 (1996) 507.

\bibitem{PDG} Review of Particle Physics, Phys. Rev. D 54 (1996) 1.

\bibitem{PPG} V.M.Aulchenko et al., e-print hep-ex/9807016, Submitted to
Phys.Rev.Lett..

\end{thebibliography}
\end{document}